\documentclass[12pt]{article}

\usepackage[utf8]{inputenc}
\usepackage[T1]{fontenc}
\usepackage{graphicx}
\usepackage{amsmath,amssymb}
\usepackage{xcolor}
\usepackage{bm}
\usepackage[margin=1in]{geometry}
\usepackage{authblk}
\usepackage{hyperref}

\title{\textbf{Tunable mesoscopic numerical model for bacterial biofilms}}

\author[1,2]{J. Martín-Roca\thanks{Corresponding author: \href{mailto:josema10@ucm.es}{josema10@ucm.es}}}
\author[3]{B. Wu-Zhang\thanks{\href{mailto:bohazhan@ucm.es}{bohazhan@ucm.es}}}
\author[4]{J. Oller}
\author[4]{J. Ramirez}
\author[1,3]{C. Valeriani\thanks{\href{mailto:cvaleriani@ucm.es}{cvaleriani@ucm.es}}}

\affil[1]{Departament de Física de la Materia Condensada, Universitat de Barcelona, C. Martı Franques 1, 08028 Barcelona, Spain}
\affil[2]{Grupo Interdisciplinar de Sistemas Complejos (GISC), Madrid.}
\affil[3]{Departamento de Estructura de la Materia, Física Térmica y Electrónica, Universidad Complutense de Madrid, Madrid, Spain.}
\affil[4]{Departamento de Ingeniería Química Industrial y del Medio Ambiente, Escuela T\'{e}cnica Superior de Ingenieros Industriales, Universidad Polit\'{e}cnica de Madrid, 28006, Madrid, Spain}

\date{\today}

\begin{document}

\maketitle

\begin{abstract}
We present a tunable mesoscale model to provide a basis for future rheological calculations of of bacterial biofilms, explicitly incorporating reversible crosslinking within the extracellular polymeric substance (EPS) matrix. Using a Dissipative Particle Dynamics framework combined with a Gillespie-inspired algorithm, bonds between polymers and bacteria dynamically form and break, capturing the intrinsically evolving nature of the network. We show that biofilm structure is governed by a competition between polymer–polymer and polymer–bacteria crosslinks, controlled by binding energy, linker availability, and bond stiffness and provide a minimal model that helps to understand the competition between both species. 
\end{abstract}


\vspace{0.5cm}

\section{\label{intro} Introduction}

\textcolor{black}{Bacterial biofilms are complex biomaterials: 
they form  on surfaces and are produced by synergistic communities of microorganisms, which are  embedded in a self-secreted matrix of extracellular polymeric substances (EPS)}
\cite{kandemir2018mechanical,picioreanu1999discrete}. EPS typically consists of polysaccharides, proteins, and extracellular DNA\cite{mann2012pseudomonas} and is surrounded by water. 

This architecture provides biofilms with superior mechanical resilience as compared to isolated bacteria, offering physical protection against external threats, which contributes to significant health issues and industrial challenges \cite{marsden2014chemotactic, attinger2012clinically, fleming2018consequences}. Mechanically, biofilms behave as complex viscoelastic materials \cite{pavlovsky2013situ,barai2016modeling,jara2021self}. They exhibit both elasticity and fluid-like rheological responses under shear, a complexity arising from compositional and topological arrangements across multiple length scales \cite{jara2021self, boudarel2018towards, gordon2017biofilms, klauck2018spatial, stewart2015artificial}. 
Thus, a detailed characterization of their mechanical properties and shear response is crucial for scientific understanding and 
technological applications, such as designing effective removal strategies \cite{aspinwall2025rigidity, persat2015mechanical}.

Given the complex structure and wide range of time and length scales characterizing biofilms, computational models are essential. Mesoscale modeling is particularly valuable, as many interesting phenomena in complex fluids occur at the mesoscale, approximately from $10$ to $10^4$ nm \cite{klotsa2019above}. In this respect, Dissipative Particle Dynamics (DPD) has emerged as a powerful coarse grained mesoscale simulation technique\cite{groot1997dissipative,espanol2017perspective}. DPD allows 
studying 
large suspensions of  particles immersed in a viscous fluid, while preserving the system's mass and momentum. Coarse-grained DPD models capture the essential topological and compositional interactions within biofilms and  predict the mechanical behavior under various stress conditions, including shear\cite{jara2021self,martin2023rheology}. Moreover, the “soft” nature of DPD interactions allows for longer integration time steps, enabling exploration of rheological time scales which are usually inaccessible to full atomistic simulations. 

The EPS matrix acts as a dynamic scaffold for cells \cite{xu2011dissipative,mukhi2022identifying}. Key structural interactions, such as EPS crosslinking (polymer-polymer or polymer-cell), contribute to the formation of transient structures which support stress, giving rise to mesoscale viscoelasticity. 
\textcolor{black}{We have recently reported mesoscale numerical models to  explore biofilm rheology \cite{jara2021self,martin2023rheology}. However, our DPD-based model relied on the assumption that the crosslinks within the polymer network were permanent
 \cite{martin2023rheology,barai2016modeling,liu2018modeling}. }

\textcolor{black}{While this numerical approach has proven successful in unraveling the mechanical properties of a {\it Pseudomonas fluorescens} biofilm, it might not be ideal when dealing with biofilms of a different structure. This is the case when dealing, for instance, with {\it Escherichia coli} biofilms. }

This discrepancy arises from the lack of dynamic crosslinking and mesoscale rearrangements \textcolor{black}{which might be observed}
under high shear forces. 

\textcolor{black}{In these cases, one can assume that }
biofilm viscoelasticity emerges from the matrix’s ability to continuously form and break 
crosslinks, allowing collective relaxation of mobile crosslinks and sliding entanglements. 
The crosslinks (or bonds) dynamics critically affect the system's mechanical properties, governing the transition from a predominantly solid-like to a more liquid-like behavior \cite{xu2011dissipative,mukhi2022identifying,stotsky2016variable}. Capturing the matrix dynamics with a reasonable detail is challenging, since parametrizing individual the dynamic of the biofilm components-including water and bonds-might quickly exceed the computational limitation.

In this work, we propose an extension of our previously established mesoscale DPD biofilm model, reported  in Ref~\cite{martin2023rheology,jara2021self}, in which we now  
incorporate dynamic bond formation and breakage within the EPS matrix. 
 \textcolor{black}{This feature is }
 implemented using a kinetic Monte Carlo (MC) method based in Gillespie-type stochastic algorithm. This approach allows the model to reproduce an "active" viscoelastic behavior and nonlinear rheology observed in 
 certain biofilms, overcoming the limitations of permanent crosslink biofilm models. Importantly, this method achieves dynamic bonding without introducing excessive molecular or algorithmic complexity, keeping the model computationally tractable and capable of predicting the rheology across extended spatiotemporal scales, as required in DPD frameworks. Studying how crosslink dynamics 
 affects  transitions in a biofilm's mechanical properties may guide future strategies for controlling  biofilm matrices, \textcolor{black}{eventually damaging it}.

This article is organized in two main parts. In Section II we describe the mesoscale numerical model used to represent bacterial biofilms, including the basic ingredients (bacteria, polymers, and water), the sample preparation and equilibration, and the key modifications introduced (as compared to our previous model\cite{martin2023rheology,jara2021self}. The latest consists in  dynamical bonds which can form and break via kinetic MC algorithm. In Section III we present the simulation results, focusing first on the number and dynamics of cross-links at equilibrium under different parameter settings, and next analyzing the system's behavior under different conditions.

\section{\label{numerical} Numerical details}

\subsection{Basic ingredients for a bacterial biofilm}

\noindent
We propose an {\it in silico} mesoscale model to study the mechanical features of viscoelastic bacterial biofilms.  According to the model developed by us previously, \cite{martin2023rheology,jara2021self}, a biofilm is represented as a mesoscale system composed of three main interacting entities: bacteria, extracellular polymeric substances (EPS), and solvent (water).

\begin{figure}[htbp]
    \centering
    \includegraphics[width=0.9\linewidth]{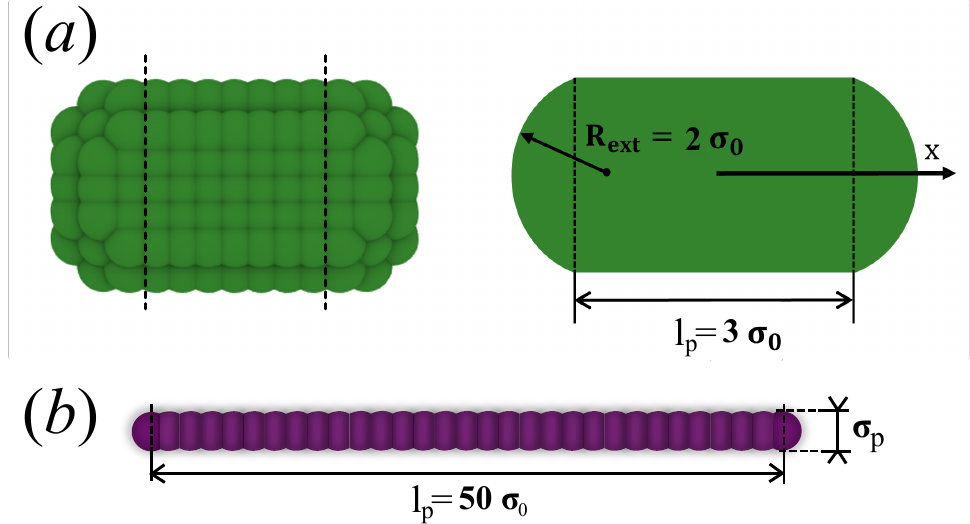}
    \caption{The illustration of bacterium in our simulations. (a) Sketch of bacteria in our simulation and illustration of the shape of our bacterium with a length of $3\sigma_0$ along the x axis and a radius of $2 \sigma_0$ of the two spherical cabs at the two ends. (b) Sketch of polymer in our simulation with the length and width used in simulations. }
    \label{fig:bacterium}
\end{figure}

\textit{Bacteria}: A single bacterium is modeled as a hollow
sphero-cylindrical shape formed by $B_b= 440$ beads (denoted as type "b"), each one with diameter $\sigma_b =0.5 \, \sigma_0$ where $\sigma_0$ is the simulation unit length. Beads are  kept together via harmonic spring interactions:

\begin{equation}
    U_{b}^{\rm bond}=K_b(r_{ij} - \sigma_b)^2,
\end{equation}
where $r_{i,j}$ is the distance between the linked beads $i$ and $j$, and $\beta \, \sigma_0^2 \, K_b=30 $ is the harmonic coupling constant, with $\beta=1/k_B T$, $k_B$  being the Boltzmann constant and $T$  the absolute temperature.  The sphero-cylinder  cell body has a length of $l_b = 3 \, \sigma_0=6 \, \sigma_b$ along the x-axis and a radius of $R_{ext} = 2 \, \sigma_0 = 4 \, \sigma_b$, as shown in Fig. \ref{fig:bacterium}. 

We set the number of bacteria in our system to be $N_b=184$, occupying $\sim 60 \%$ of the total volume (taking into account the excluded volume interactions explained below). \textcolor{black}{The volume fraction is calculated as $f_b = N_bV_b/V_{box}$, where $V_b= \frac{4}{3} \pi (R_\text{ext}+R_{vol})^3 + \pi (R_\text{ext}+R_{vol})^2 \, l_b$ is the volume of each bacteria with the excluded volume radius $R_{vol}=0.5\, \sigma_0$} and $V_{box}$ is the volume of the simulation box. 

When preparing the initial configuration,  bacteria are randomly inserted in the simulation box. 

\textit{Extracellular Polymeric Substances (EPS)}: The EPS polymer matrix consists of $N_p = 80$ polymers.  A single polymer is modeled as a freely-joint linear chain consisting of $B_p = 100$ beads, connected by a harmonic potential 

\begin{equation}
    U_{p}^{\rm bond}=K_p(r_{ij} - \sigma_p)^2
\end{equation}
where $r_{i,j}$ is the distance between two consecutive beads $i$ and $j$, $\sigma_p$ the equilibrium distance, and $K_p$ the coupling constant $\beta \, \sigma_0^2 \,  K_p=30$. Each bead (namely, type "p") in the polymer chain has a radius of $\sigma_p = 0.5  \, \sigma_0$, indicating a contour length of $l_p = 50 \sigma_0$. 

$N_p = 80$ of such polymers together occupy a total volume of $N_p \, V_p$. The occupied volume of a single polymer \textcolor{black}{is calculated considering the volume of each bead and the total length of the polymer, $V_p=l_p \, \pi \left(\frac{1}{2}  \sigma_p +R_{vol} \right)^2$}. Therefore, $V_p$ for a single polymer has a value of $35.63 \sigma_0^3$. Furthermore, with the relation $f_p = N_pV_p/V_{box}$, we  obtain $f_p = 2.40\%$.

When preparing the initial configuration, similarly to bacteria,  polymer chains are placed randomly in the simulation box. 

\textit{Water solution}: Water is represented by $N_{\rm solv} = 35000$ solvent beads of type $s$. Each bead has a diameter of $\sigma_s=\sigma_0$, and it effectively models an aggregate of several water molecules with one single soft particle.   

\textcolor{black}{When preparing the initial configuration,  water beads are randomly inserted in the simulation box}. 

\noindent
\textit{Cross-links(CL)}: In our system, we allow cross-links (or bonds) to form between two polymer beads (type $p$), and between a polymer bead and a bacteria bead (type $b$). Note that the cross-links between polymer beads are only allowed between two polymer beads belong to different chains (a.k.a. intra-polymer interactions are not allowed, see Fig.\ref{fig:CL-equilibrium}c.). The cross-linking (CL) interactions are modeled via a harmonic potential:
\begin{equation}
    U_{CL}^{\rm bond}=K_{CL}(r_{ij} - \sigma_{CL})^2
\end{equation}
where $K_{CL}$ is a  tuning parameter, with    $\beta \, \sigma_0^2 \, K_{CL}=30$. $\sigma_{CL}$ is the equilibrium length of a cross-links, and takes the value of  $\sigma_{CL}= 0.5 \, \sigma_0$. 

\subsection{How to prepare the initial  configuration} \label{sec:initialconf}

\noindent
The initial configurations consists of  a mesoscopic system of interacting beads representing bacteria, polymers and water, embedded in a  $32 \sigma_0\times32\sigma_0\times 32\sigma_0$ box with periodic boundary conditions. As described above, all  entities in our system are modeled as a group of beads  some of which can be connected with each other by \textcolor{black}{reversible} bonds (between bacteria and polymer). The time evolution of the beads is achieved by Dissipative Particle Dynamics (DPD), simulated via the open source LAMMPS numerical package\cite{lammps}.  An implementation of the permanent code can be found in \cite{martin2023rheology}.

All relevant binding interactions are encoded in the mesoscale simulation domain: polymer-polymer (pp), polymer-bacteria (pb), polymer-solvent (ps), bacteria-bacteria (bb), bacteria-solvent (bs), and solvent-solvent (ss). 
All beads  evolve according to DPD, in which the total force between pairs of particles $i$ and $j$ can be expressed as the sum of a conservative force $\vec{F}^C_{ij}$, a dissipative force $\vec{F}^D_{ij}$ and a random force $\vec{F}^R_{ij}$.
The sum of the three forces captures not only the long-range correlations induced by  hydrodynamics but also  thermal fluctuations.
The conservative force is given by 
\begin{equation} \label{eqn:F_c}
\vec{F}^C_{ij} = A_{\alpha,\beta}w(r_{ij})\hat{r}_{i,j} \quad \text{for } r_{ij} < r_c,
\end{equation}
 where $r_c$ is a cutoff distance beyond  which all  terms vanish;   $\hat{r}_{ij}$ is the unit vector along the direction of $\vec{r}_i - \vec{r}_j$, $A_{\alpha,\beta}$ is the conservative interaction coefficient for particle types $\alpha$ and $\beta$. The dissipative force is given by 
\begin{equation}
\vec{F}^D_{ij} = -\gamma w^2(r_{ij})(\hat{r}_{ij} \cdot \vec{v}_{ij})\hat{r}_{ij},
\end{equation}
where $\vec{v}_{ij}$ is the vector difference between the velocities of the particles $i$, and the velocity $\vec{v}_j$ of the particle $j$. Here, $\gamma$ is the friction coefficient. 
The random force can be calculated as 
\begin{equation}
\vec{F}^R_{ij} = w(r_{ij}) \left( \frac{2k_B T \gamma}{\Delta t} \right)^{1/2} \Theta \,  \hat{r}_{ij},
\end{equation}
with  $ w(r_{ij}) = 1-r_{ij}/r_c$ a weighting factor varying from 0 to 1. The quantity $\Theta$ is a  normally distributed  Gaussian variable; $\Delta t$ is the integration time.

Using non-dimensional units for all  quantities, we set $r_c = \sigma_0=1$ for all DPD interactions as the unit  distance, the  reduced temperature $k_B T=1$ as the unit energy. The masses of bacteria ($m_b$), polymer($m_p$) and solvent ($m_s$) are set to $m_0=1$, resulting in the time scale $\tau_0 = \sqrt{\sigma_0^2\, m_0/k_BT}=1$ as the unit of time. 

Therefore, the friction coefficient is set as $\gamma= 4.5\, m_0/\tau_0$, and time step $\Delta t=0.05 \, \tau_0$. As explained in Ref. \cite{martin2023rheology}, to ensure that bacteria and polymers \textcolor{black}{are properly surroundder by solvent}, we choose the conservative interaction coefficient (equation \ref{eqn:F_c}) of the interactions between solvent particles (type "s") and others to be smaller than the value of the same coefficient for any other pair of interactions. This implies that we choose $A_{s,s} = A_{s,b} = A_{s,p} = 25 \, k_BT/\sigma_0 $ and $A_{b,b} = A_{p,p} = A_{p,b} = 30 \, \, k_BT/\sigma_0$.

\subsection{Novelty with respect to our previous mesoscale model} \label{sec:gilespi}

Based on the mesoscale model we reported in ~\cite{martin2023rheology,jara2021self}, we have tailored a novel model which allows to  describe 
biofilms characterised by viscoelastic features that cannot be captured within a framework of permanent crosslinks. 
Unlike the previous approach, in which crosslinks were treated as static constraints,  crosslinking (CL) interactions are now explicitly considered as dynamic and reversible. This modification allows bonds between particles to be formed and broken during a simulation run. Even though  such events occur stochastically, the system converges towards a steady state in which the average number of crosslinks fluctuates around a well-defined equilibrium value. Therefore, the main aim of the present model is  to explore the parameter space  governing the balance between bond creation and breakage, thus enabling us to tune the density and lifetime of crosslinks according to experimentally relevant conditions.  

The dynamic crosslinking is governed by a kinetic Monte Carlo algorithm based on the Gillespie formulation of chemical master equations \cite{gillespie1976general, gillespie1977exact}. In practice, the Gillespie update is attempted every $\tau_G=2\tau_0$, corresponding to 40 DPD time steps in our simulations, during which the probabilities of forming and breaking bonds are evaluated.  This choice corresponds to a  sufficiently large beads relaxation time, to avoid instantaneous bond creation and breakage  (leading to unrealistic dynamics),  and to a sufficiently small value of $\tau_G$, to capture the bond dynamics within the simulated timescale and reach the equilibrium state.  The choice of the specific value of $\tau_G$ relies on preliminary runs showing a mild dependence of $\tau_G$ on the number of crosslinks.  Importantly, this update frequency does not alter the intrinsic kinetics, which are solely controlled by the energetic parameters of the model (see Appendix \ref{append:tauMC}). 

The essential input needed  is a set of thermodynamic and kinetic parameters which translate microscopic energetic considerations into statistical event probabilities. Two quantities play a central role. The first is the bonding energy, $E_e$, which reflects the  effective potential-well minimum associated with a crosslink. \textcolor{black}{This parameter does not represent  the stiffness of the harmonic bond, $E_e$ primarily represents the equilibrium energy of the bonded state: larger values of $E_e$ better stabilise the bonded state, thereby shifting the balance towards a larger number of persistent links.}

The second parameter is the activation energy, $E_a$, which embodies the kinetic barrier for both bond-formation and breakage. In analogy with the Arrhenius kinetics, the factor $\exp(-E_a/\text{R}T)$ sets the intrinsic probability per unit time of overcoming this barrier, thereby modulating the timescale of the stochastic dynamics without directly fixing the equilibrium state (see Appendix). In the last expression $\text{R}$ is the ideal gas constant considered  as a fundamental unit for $E_e$ and $E_a$ ($\text{R}=1$).

\textcolor{black}{Inspired by the $\tau$-leaping variant of Gillespie’s stochastic simulation algorithm \cite{gillespie1976general, gillespie1977exact}, we tailor a procedure which ensures that the discrete and random nature of reaction events is preserved while allowing multiple events to occur within a finite timestep.}
At each update, the number of events is drawn from a Poisson distribution whose mean is given by the product of the  time window $\tau_G$ and the total reaction propensity ($\lambda$), namely
\begin{equation}
n_{\mathrm{events}} \sim P(\lambda \cdot \tau_G).
\end{equation}

The propensity $\lambda$ quantifies the instantaneous likelihood that a given microscopic reaction will occur, and it depends on whether a bond has been  formed or broken. 
For {\it bond creation} between two eligible particles $i$, the propensity is fixed as
\begin{equation}
\lambda^c_i = \exp\!\left[-\frac{E_a}{\text{R}T}\right],
\end{equation}
so that all pairs of available pairs share the same kinetic barrier. In contrast, for {\it bond breakage} the algorithm accounts not only for the intrinsic barrier $E_a$, but also for the instantaneous mechanical state of the bond. This is achieved by introducing a shifted bond potential, $U_{bs}(r_i,E_e)$, which measures the energy cost of stretching a bond relative to its equilibrium distance $r_0$. The potential is then shifted  
so that the minimum potential is zero:

\begin{equation}
U_{bs}(r_i,E_e) = U_b(r_i) - U_b(r_0) - E_e,
\end{equation}
\textcolor{black}{The elastic contribution is shifted so that its minimum is zero, while the bonded state is stabilized by an additional energy Ee.} Therefore, propensity for bond breakage is 
\begin{equation}
\lambda^b_i = \exp\!\left[-\frac{E_a - U_{bs}(r_i,E_e)}{\text{R}T}\right],
\label{eq:break}
\end{equation}
which encapsulates the fact that \textcolor{black}{stretched or weakened bonds are more prone to rupture then equilibrium or compressed bonds}. In this way, the breaking rate emerges from a subtle interplay between thermal fluctuations, the imposed activation barrier, and the current deformation energy stored in the bond.  \textcolor{black}{Note that this algorithm has been tailored to naturally allow for simulating system in which non-equilibrium deformations take place.}

In addition to this energetic criterion, the implementation includes two geometric cutoffs that guarantee physical plausibility. On the one hand, new bonds can only be created if the distance between reactive partners is smaller than $r_{\max}$, whereas bonds shorter than $r_{\min}$ are excluded from rupture attempts. \textcolor{black}{ In this particular case we decided to use $r_{ \min}=0$ and $r_{\max}=\sigma_0$. On the other hand, any bond whose elastic bond length overcomes a predefined threshold $R_{\max}$ is automatically broken, bypassing the stochastic criterion. This rule prevents the system from sustaining unphysical, highly stretched bonds that would otherwise dominate the dynamics. In practice, $R_{\max}$ is chosen such that bonds close to this limit correspond to highly unstable configurations, so that rupture is always enforced. In our simulations, we set  $R_{\max}=\sigma_0$. }

The total reaction propensities for creation and breaking, $\lambda^c$ and $\lambda^b$, are then obtained by summing over all candidate pairs or existing bonds, respectively
\begin{equation}
    \lambda^c = \sum_i \lambda^c_i \quad \mathrm{and} \quad \lambda^b = \sum_i \lambda^b_i
\end{equation}
The stochastic sampling of events based on these propensities ensures that the simulation faithfully reproduces the probabilistic character of crosslink dynamics, with the kinetic MC algorithm determining not only how many events occur during $\tau_G$ but also which specific pairs are involved, \textcolor{black}{proportional to their propensities.} From a physical perspective, this framework couples the kinetics of reversible bonding to the thermodynamic bias imposed by $E_e$, thereby allowing the emergent crosslink density and bond lifetime. The inclusion of $r_{\max}$, $r_{\min}$, and $R_{\max}$ makes the dynamics additionally sensitive to geometric constraints, ensuring that the simulated crosslinks behave as physically consistent harmonic bonds.

\subsection{How to prepare a equilibrium configuration} \label{sec:steadystate}

Finally, to achieve a steady state configuration, first we run an equilibration run for $10^3 \tau_0$. The activation energy is  chosen as $E_{a,pb}=E_{a,pp}=E_a=4$, which controls the bond creation/destruction. This value has been set from preliminary simulations  showing a not relevant dependency of $E_a$ on the number of CLs for our propose (see Appendix).

After reaching steady state, we run long enough  production runs to allow the system to reach equilibrium, the length of the equilibration is set to $10^{5} \, \tau_0$, so the number of bond creation/destruction events is  $N_G=10^5$. Finally, the steady state configurations were used for  production runs  for a total time of $N_G \, \tau_G=2\cdot 10^{5} \, \tau_0$.  

\begin{figure}[htbp]
    \centering
    \includegraphics[width=1\linewidth]{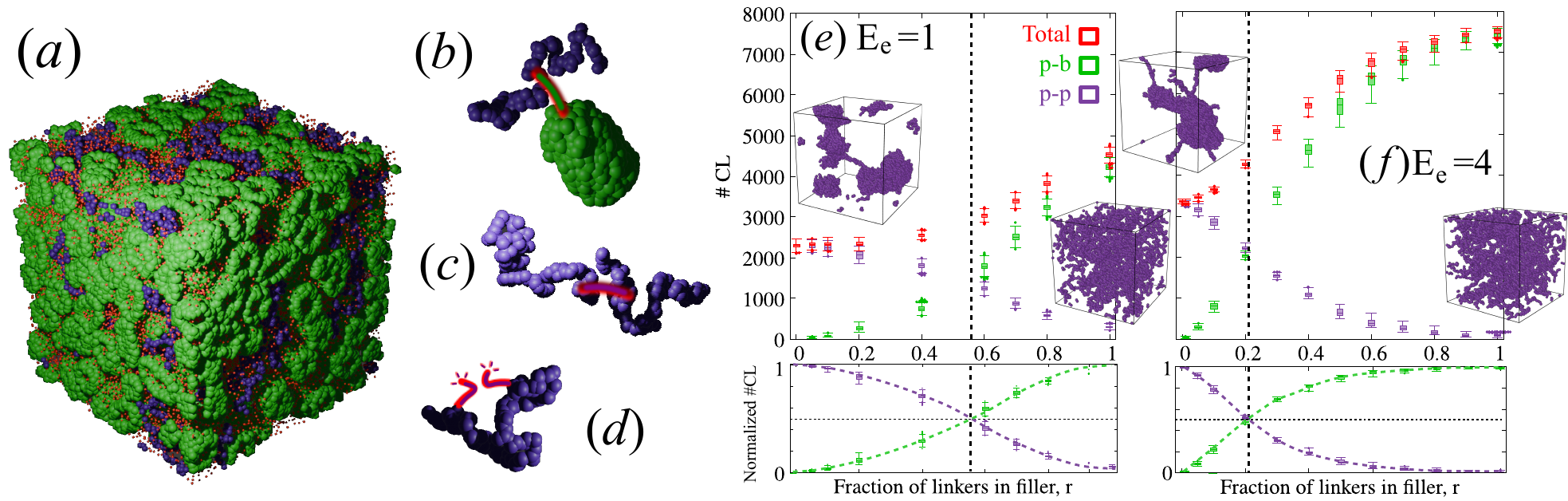}
    \caption{(a) Snapshot of the system under study in the steady-state.(b) Cross-Link between polymer-bacteria. (c) Cross-Link between polymer-polymer. (d) Cross-Link within the same polymer is not considered in this model. (e-f) Number of Cross-Links for $E_{e,pp}=E_{e,pb}=1$ (e) and $E_{e,pp}=E_{e,pb}=4$ (f) with $E_{a,pp}=E_{a,pb}=4$ as a function of the fraction of bacteria beads linkers, $r$, for the different species: polymer-bacteria (green), polymer-polymer (purple) and the total number of CL (red). At the bottom of each panel we represent the value of the fraction of cross-links respect to the total number of cross-links on each case. We present some snapshots of the system for specific values of $r$ on each panel, to show the transition between different networks depending on the predominant type of CL.  }
    \label{fig:CL-equilibrium}
\end{figure}

\textcolor{black}{Figure\ref{fig:CL-equilibrium}.a. reports a snapshot of the steady state biofilm, which includes bacteria (green), polymers (purple) and solvent (red), as previously described. Panels (b) and (c) show schematic representations of the possible cross-links (CLs) in the system, namely polymer–bacteria cross-links (panel b) and polymer–polymer cross-links (panel c). These bonds are generated by means of the MC algorithm described above.}

\section{RESULTS} \label{sec:results}

To properly characterise the numerical model,  we are going to tune the following controlling parameters for the number of crosslinks CL:  the polymer-polymer bonding energy , $E_{e,pp}$; the polymer-bacteria bonding energy, $E_{e,pb}$; the sticky area fraction randomly place in the surfase (\textcolor{black}{fraction of bacteria beads that can perform CL) $r$;}  the CL elastic constant, $K_{CL}$. \textcolor{black}{All bacteria beads that can perform CL are placed randomly along the bacteria}. This will provide us with a controllable way to unravel the relevance of each parameter in the creation or breakage of crosslinks within the biofilm. Note that the distance between pairs of particles is indicated as $r_{ij}$ and not $r$. 

\subsection{Dependence on the sticky area fraction}

While polymers are allowed to form bonds with other polymers via any of their beads, we assume that only a fraction $r$ of the beads located on the bacterial surface can form bonds with polymers: this is what we call sticky area fraction. 
\textcolor{black}{We start by computing the number of CL of each specie present in the system (polymners and bacteria) to characterize the competition between both.}

Panel (e) and (f) in Figure\ref{fig:CL-equilibrium} display the number of CLs as a function of the parameter $r$ for two different binding energies, $E_e=1$ (panel e) and $E_e=4$ (panel f). A key feature emerging from Fig.~\ref{fig:CL-equilibrium}.e and f is the existence of a clear crossover between two distinct cross-linking regimes.  At low values of the active surface fraction $r$, the system is dominated by polymer-polymer (p-p, purple curves in the bottom panel) cross-links, whereas at large $r$ polymer-bacteria (p-b, green curves in the bottom panel) cross-links become dominant.  This crossover is not only reflected in the number of bonds, but also in the spatial organization of the system, as illustrated by the snapshots insets included in Fig.~\ref{fig:CL-equilibrium}.e and f.  At low $r$, polymers form a connected network via mutual linking, whereas at high $r$ polymers become increasingly anchored to the bacterial surfaces, leading to a qualitatively different network topology.

This behavior can be understood as a competition between two bonding channels that share the same pool of polymer binding sites. Each polymer bead can either connect to another polymer or to a bacterial site, but not to both simultaneously.  Therefore, increasing the number of available bacterial linkers (through $r$) effectively redirects polymer connectivity from polymer-polymer linking to polymer-bacteria anchoring.  This competition introduces a non-linear redistribution of cross-links between the two species.

In order to rationalize this behavior, it is useful to consider the fraction of cross-links of each type, rather than their absolute number. We define $x(r) = \frac{N_{pp}}{N_{pp} + N_{pb}}$, and $y(r) = \frac{N_{pb}}{N_{pp} + N_{pb}}$, which by construction satisfy $x+y=1$. These quantities directly capture the competition between bonding channels and provide a natural way to compare systems with different total numbers of cross-links. 

Remarkably, the simulation data can be accurately described by a simple phenomenological model of the form
\begin{align}
x(r) = \frac{w_{pp} (r_0-r)^{\alpha}}{w_{pp} (r_0-r)^{\alpha} + w_{pb} r^{\beta}}, \\
y(r) = \frac{w_{pb} r^{\beta}}{w_{pp} (r_0-r)^{\alpha} + w_{pb} r^{\beta}},
\label{eq:fraction_model}
\end{align}
where $w_{pp} \sim e^{E_{e,pp}/RT}$ and $w_{pb} \sim e^{E_{e,pb}/RT}$ encode the energetic preference for each type of bond.
\textcolor{black}{The parameter $r_0$ contains the information of the saturation of CL at $r=1$ (maximum number of bacteria linkers). Finally, $\alpha$ and $\beta$ are effective exponents  accounting for the combinatorial and geometrical constraints of the system. Eq.(12) and (\ref{eq:fraction_model}) can be interpreted as a competition between two statistical weights.  The factor $(r_0-r)^{\alpha}$ reflects the decreasing availability of polymer-polymer partners as more bacterial sites become active; while $r^{\beta}$ represents the increasing number of accessible polymer-bacteria bonding configurations.}
We should note that for $r=0$ only p-p CL can take place in the system, i.e. $x(0)=1$, $y(0)=0$. However, for $r=1$, not necessarily only p-b CL are present in the system:  in general, $x(r=1) \neq 0$ and $y(r=1) \neq 0$.

The crossover point $r^*$ is naturally defined by the condition $x=y$, which yields
\begin{equation}
\frac{(r_0-r^*)^{\alpha}}{(r^*)^{\beta}} = \frac{w_{pb}}{w_{pp}} 
= \exp\!\left(\frac{E_{e,pb} - E_{e,pp}}{RT}\right),
\label{eq:rstar}
\end{equation}
This equation connects the location of the crossover directly to the microscopic interaction energies. This result explains the shift observed in Fig.~\ref{fig:CL-equilibrium}: increasing $E_e$ enhances the statistical weight of bonds, but the relative balance between both species determines where the transition occurs.

\begin{figure}[htbp]
    \centering
    \includegraphics[width=0.9\linewidth]{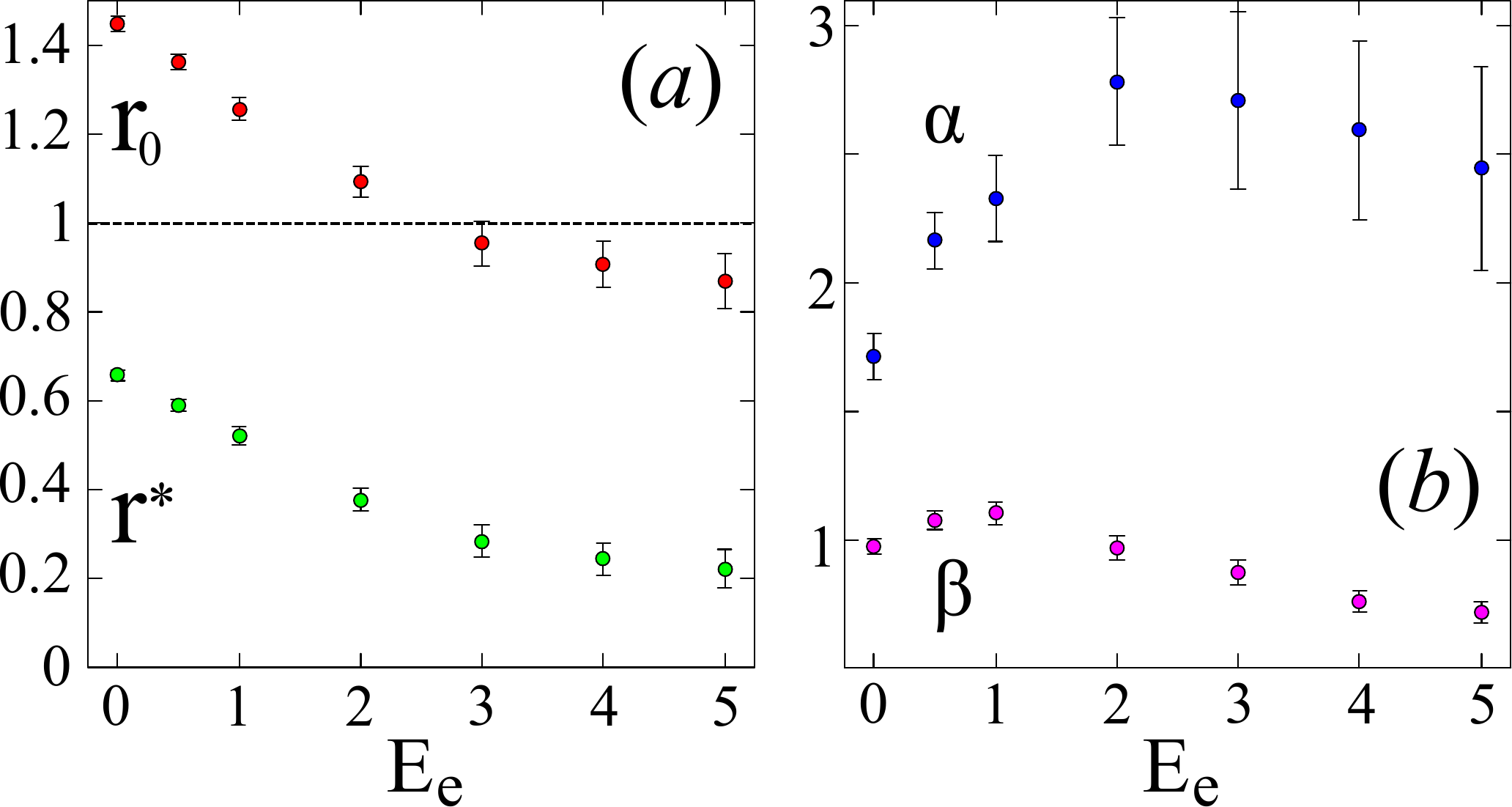}
    \caption{Parameters obtained from  fitting  Eq. \ref{eq:fraction_model} of normalized number of CL of each specie as a function of $r$ for different  $E_e$. (a) Value of $r$ for the cross-over between CL for p-p and for p-b, $r^*$ (green) and saturation $r$ value for which $x(r_0)\approx 0$ and $y(r_0) \approx 1$ (red). (b) Exponents $\alpha$ (blue) and $\beta$ (pink) as function of $E_e$.}
    \label{fig:CL-theory}
\end{figure}

\textcolor{black}{In Fig.~\ref{fig:CL-theory}(a) we present the value of $r^*$ for which the cross-over between both species of CL occurs. For that value the transition between both network structures takes place.} 

\textcolor{black}{
We observe that     $r^*$ decreases with $E_e$, indicating that the cross-over happens for a small amount of linkers in a bacteria with higher bonding energy. }

On the other hand,  $r_0$ is larger than one for given values of $E_e$. For those values, the saturation of each specie is not reached for any $r$ (since $0 \leq r \leq 1$). Clearly, $r^*$ and $r_0$ are correlated via Eq. \ref{eq:rstar}.

A deeper insight  is obtained by analyzing the fitted values of the exponents $\alpha$ and $\beta$, shown in Fig.~\ref{fig:CL-theory}(b):    $\alpha > 2$ while $\beta \lesssim 1$; moreover,  $\alpha$ exhibits a non-monotonic dependence on the binding energy. This asymmetry reflects the fundamentally different nature of the two bonding processes.  Polymer-polymer bonding requires the simultaneous availability of two compatible polymer sites within the interaction-range, making it highly sensitive to crowding and network structure.  In contrast, polymer-bacteria bonding involves a polymer and a comparatively static surface site; thus,  it is geometrically less constrained. 

From a physical perspective, $\alpha$ can be interpreted as an effective connectivity exponent that measures how rapidly the availability of polymer-polymer bonding configurations decreases as the system becomes saturated with polymer-bacteria links.  The fact that $\alpha$ is larger than unity indicates a strong cooperativity or excluded-volume effects, while its non-monotonic dependence on $E_e$ suggests a competition between bond stabilization and structural heterogeneity.

To summarize, the crossover between the two bonding regimes can be connected to classical theories of gelation and percolation.  In the present system, p-p bonds contribute to the formation of a polymer network, while p-b bonds effectively anchor polymers to a secondary network formed by the bacterial phase. The transition at $r^*$ therefore corresponds to a change in the dominant connectivity channel, from a polymer-driven percolating network at low $r$ to a bacteria-mediated network at high $r$. Importantly, this is not a thermodynamic phase transition in the strict sense, but rather a structural transition between two different network organizations. 

\subsection{Dependence on the cross-link stiffness}

\begin{figure}[htbp]
    \centering
    \includegraphics[width=0.9\linewidth]{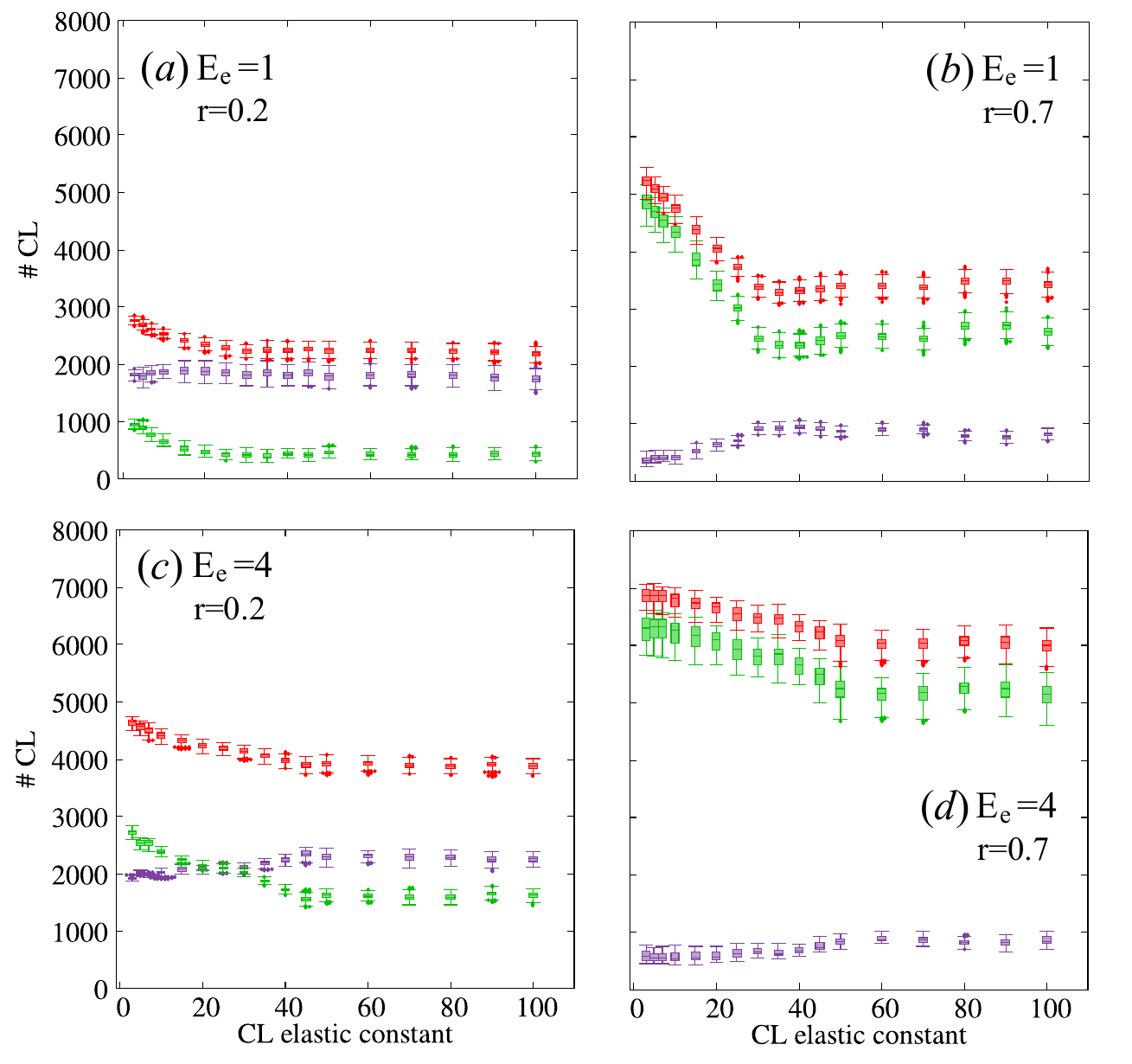}
    \caption{Number of CL for $E_a=4$, $\tau_{G}=2$, $k_{pp}=30$ as a function of CL elastic constant ($K_{CL}$) using different combination of bonding energy and fraction of bacteria linkers: (a) $E_e=1$, $r=0.2$;  (b)$E_e=1$, $r=0.7$; (c)$E_e=4$, $r=0.2$; and (d)$E_e=4$, $r=0.7$.  In all panels, the number of CL for polymer-bacteria  are in green, the number of CL for polymer-polymer  in purple and the total number of CL in red.} 
    \label{fig:CL-equilibrium-kCL}
\end{figure}

In Fig.~\ref{fig:CL-equilibrium-kCL}, we analyze the effect of the cross-link elastic constant $K_{CL}$ on the equilibrium number and distribution of cross-links (CLs).  Unlike the parameters previously discussed, which primarily control the thermodynamic balance between bonding species, $K_{CL}$ directly affects the mechanical stability of the bonds and therefore their lifetime within the stochastic dynamics.

We observe a common feature across all explored conditions: the total number of cross-links (in red)  decreases with increasing $K_{CL}$ before reaching a plateau at large stiffness.  This behavior can be directly traced back to the bond-breaking propensity introduced in Eq.~\ref{eq:break}, $\lambda^b \sim \exp\left[U_{CL}/RT\right]$. For a harmonic bond, the elastic contribution scales as $U_{CL} \sim K_{CL}(r_i-\sigma_b)^2$, so that the brekage rate increases exponentially with both bond extension and stiffness.  As a consequence, even moderate fluctuations around the equilibrium bond length lead to a significant enhancement of bond breaking when $K_{CL}$ is large.

Although this mechanism explains the overall decrease in the number of CLs, a more subtle and important effect emerges when the two bonding species are considered separately. As $K_{CL}$ increases, the system exhibits a systematic redistribution of cross-links, i.e. polymer-polymer  bonds (p-p, purple curves in Fig.\ref{fig:CL-equilibrium-kCL}) increase, whereas polymer-bacteria  bonds (p-b, green curves in Fig.\ref{fig:CL-equilibrium-kCL}) decrease.  This redistribution is non-trivial and reveals that bond stiffness does not affect both species equally, but instead biases the competition between the two network-forming channels.

\noindent
\textcolor{black}{ This behavior can be understood as a consequence of geometrical constraints and local confinement. Polymer-bacteria bonds connect a flexible polymer segment to a relatively rigid surface. When such a bond becomes stiff, the attached polymer segment loses configurational freedom and its fluctuations become more constrained. However, confinement alone does not necessarily reduce the probability of forming additional contacts with the same bacteria. A more robust effect arises from geometric compatibility when the formation and maintenance of a stiff bond require the two partners to remain close to the preferred bond distance imposed by the harmonic potential. While polymer segments can rearrange cooperatively to satisfy this constraint, bacterial surface beads are embedded in a rigid structure and possess fewer configurational degrees of freedom. As $K_{CL}$ increases, deviations from the optimal bond geometry become increasingly penalized, enhancing bond rupture and reducing the stability of polymer-bacteria bonds relative to polymer-polymer bonds.}

Together, these mechanisms explain the non-monotonic behavior of the total number of CLs. In an intermediate range of $K_{CL}$, the loss of p-b bonds dominates over the gain of p-p bonds, leading to a net decrease in connectivity. At sufficiently large $K_{CL}$, however, the system reaches a connectivity-limited regime in which the number of bonds is controlled solely by the availability of binding sites and the thermodynamic affinity set by $E_e$.  In this limit, a further increase in stiffness does not significantly alter the number of available bonds, leading to the observed plateau.

This redistribution can be interpreted within the same framework introduced in the previous section.  Increasing $K_{CL}$ effectively modifies the statistical weights of the bonding channels, i. e. p-b bonds become less favorable due to geometric incompatibility and enhanced breakage, while p-p bonds are comparatively stabilized.  In the language of the phenomenological model, this corresponds to a stiffness-dependent renormalization of the weights $w_{pp}$ and $w_{pb}$, and therefore to a shift in the effective balance between the two competing networks.

\begin{figure}[htbp]
    \centering
    \includegraphics[width=0.9\linewidth]{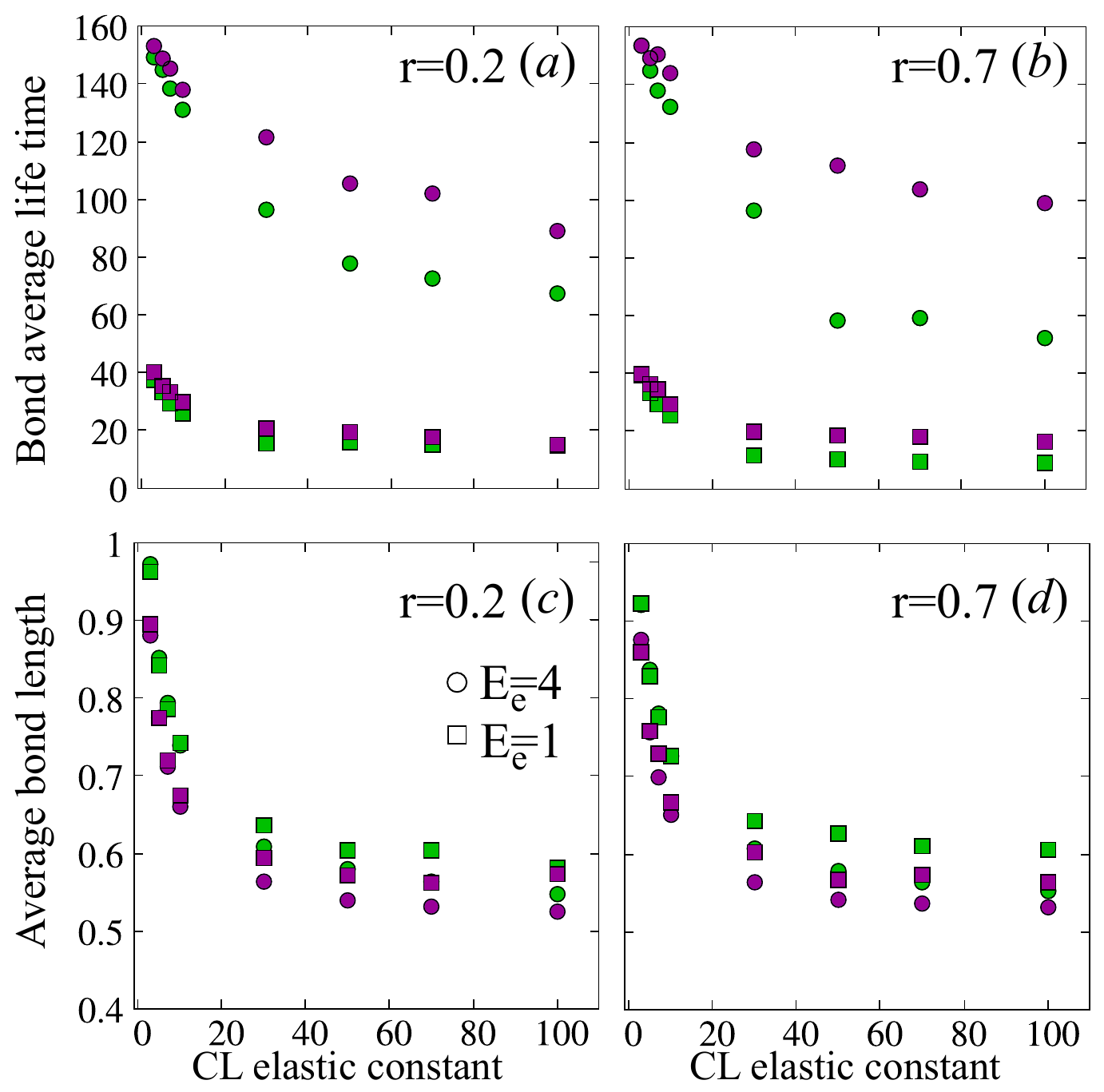}
    \caption{Average bond life time (top) and bond length (bottom) of p-p (purple) and p-b (green) bonds for $E_a=4$, $\tau_{G}=2$, $k_{pp}=30$ as a function of CL elastic constant ($K_{CL}$) with different combination of bonding energy and fraction of bacteria linkers: (a) $E_e=1$, $r=0.2$;  (b)$E_e=1$, $r=0.7$; (c)$E_e=4$, $r=0.2$; and (d)$E_e=4$, $r=0.7$.  }
    \label{fig:life-length}
\end{figure}

To further elucidate the microscopic origin of this effect, we analyze in Fig.~\ref{fig:life-length} both the bond lifetime and the bond length distribution for each species. The average bond lifetime was computed by tracking the temporal persistence of each individual crosslink throughout the simulation. For every saved frame, all polymer–polymer (p–p) and polymer–bacteria (p–b) bonds were identified from the bond files, and each bond was uniquely labeled by the pair of particle IDs involved. A bond was considered to survive as long as the same pair remained connected in consecutive frames. When a bond disappeared, its lifetime was recorded as the total elapsed simulation time between its formation and rupture. Repeating this procedure for all bonds generated a lifetime distribution for each bond species, from which the mean bond lifetime was obtained by direct averaging. Simulation steps were converted into physical simulation time using the integration timestep. 

These quantities provide a direct insight into how the stochastic rules of the algorithm translate into the observed macroscopic behavior.

The results in panels (a) and (b) show that polymer-bacteria bonds (in green) systematically exhibit shorter lifetimes than polymer-polymer bonds (in purple), particularly at large $K_{CL}$.  This is consistent with their reduced geometric compatibility, i. e. even small deviations from the preferred bond length result in a large elastic penalty, which significantly enhances their breakage probability.  In contrast, polymer-polymer bonds (in purple) can redistribute deformation more efficiently, allowing them to remain stable for longer times.

The analysis of bond lengths in panels (c) and (d) reveals a complementary picture. At large $K_{CL}$, both types of bonds fluctuate around the equilibrium distance, but their lifetimes differ markedly, reflecting their distinct stability. At small $K_{CL}$, however, bonds become highly deformable and can reach extensions close to the imposed cutoff $G_{\max}$. Despite this large deformation, the associated elastic energy remains small due to the low stiffness, and therefore the breakage propensity is only weakly affected. As a result, bonds can persist for long times even in highly stretched configurations.

\subsection{Dependence on the binding asymmetry}

\begin{figure}[htbp]
    \centering
    \includegraphics[width=0.9\linewidth]{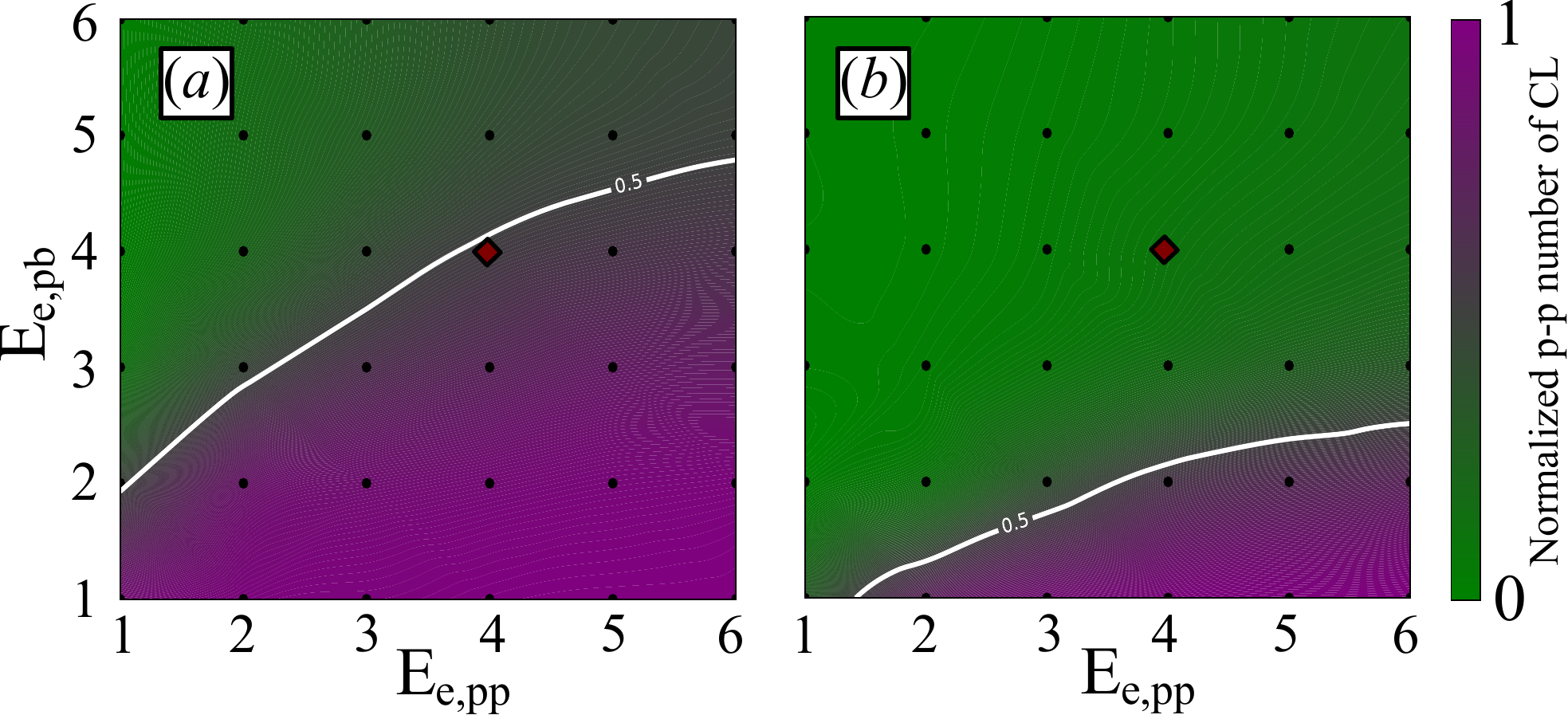}
    \caption{Normalized number of p-p CL for $E_a=4$, $\tau_{G}=2$, $k_{pp}=30$, $K_{CL}=30$ using different combination of bonding energy for each bond specie and fraction of bacteria linkers for (a) $r=0.2$ and   (b) $r=0.7$. Points represent the measured values in simulations which have been interpolated to construct the continuous colored phase diagram.}
    \label{fig:CL-Energy}
\end{figure}

In Fig.~\ref{fig:CL-Energy} we decouple the binding energies of polymer-polymer ($E_{e,pp}$) and polymer-bacteria ($E_{e,pb}$) cross-links in order to disentangle their relative contributions to network formation.  While previous results showed that increasing a single global binding energy $E_e$ enhances the total number of CLs, introducing an asymmetry between $E_{e,pp}$ and $E_{e,pb}$ reveals a much richer and more selective redistribution mechanism.

The most important observation is that the system's response is not symmetric with respect to the two bonding channels. Increasing $E_{e,pb}$ strongly enhances the formation of polymer-bacteria bonds, rapidly suppressing polymer-polymer connectivity even at relatively low values of $r$ (panel a).  In contrast, increasing $E_{e,pp}$  promotes polymer-polymer bonding, but does not completely suppress polymer-bacteria links, especially at large $r$ (panel b). This asymmetry indicates that the two types of bonds play intrinsically different structural roles.

This behavior can be understood in terms of the competition between bonding channels introduced in previous sections. As we have shown, the relative abundance of each species is controlled by their statistical weights. Therefore, the balance between both populations is governed by the ratio
\begin{equation}
\frac{w_{pb}}{w_{pp}} = \exp\!\left(\frac{E_{e,pb} - E_{e,pp}}{RT}\right),
\end{equation}
which directly controls the crossover between bonding regimes.

Within the phenomenological framework proposed above, this imbalance shifts the crossover point $r^*$ according to Eq. \ref{eq:rstar}, explicitly showing  that the transition between the network structures is governed by the difference in the binding energies, rather than their absolute values. As a consequence, increasing $E_{e,pb}$ shifts the crossover towards lower $r$, while increasing $E_{e,pp}$ shifts it towards higher $r$, in complete agreement with the trends observed in Fig.~\ref{fig:CL-Energy}.

From a structural perspective, this asymmetry reflects the different connectivity roles of the two bonding species. As shown in Fig. \ref{fig:CL-equilibrium}, polymer-polymer bonds contribute to the formation of an extended, flexible network in which connectivity is distributed throughout the polymer phase. In contrast, polymer-bacteria bonds act as anchoring points that localize polymers onto a relatively rigid connected clusters. When $E_{e,pb}$ is large, this anchoring mechanism dominates and rapidly captures polymer segments, effectively suppressing the formation of extended polymer networks. Conversely, increasing $E_{e,pp}$ strengthens the polymer network but does not eliminate polymer-bacteria bonding, since the latter is still geometrically favored at large $r$.

This mechanism explains why the system response is inherently asymmetric, i.e. polymer-bacteria bonds compete not only energetically but also geometrically, as they rely on a distinct pool of binding sites whose availability is controlled by $r$. As a result, the interplay between energy asymmetry and linker availability leads to a continuous but highly non-linear redistribution of connectivity between the two networks.

\section{Conclusions}

In this work, we have developed a tunable mesoscale model for bacterial biofilms which explicitly incorporates reversible and stochastic crosslinking within the extracellular polymeric substance (EPS) matrix. By extending previous fixed-crosslinks DPD approaches \cite{martin2023rheology,jara2021self}, we introduced a kinetic Monte Carlo algorithm which allows bonds between polymers and between polymers and bacteria to dynamically form and break. This feature enables the system to reach a non-trivial steady state characterized by fluctuating and well-defined crosslinks population, providing a physically consistent framework to study biofilm mechanics beyond permanent network approximations.

A central result of this work is the identification of a competition mechanism between two distinct bonding channels: polymer-polymer and polymer-bacteria crosslinks. These two species compete for a common pool of polymer binding sites, while polymer-bacteria bonds are additionally limited by the finite number of active sites on the bacterial surface, controlled by the parameter $r$. This competition leads to a continuous but well-defined crossover between two structurally distinct network regimes: a polymer-dominated network at low $r$, and a bacteria network with branches of polymers at high $r$.

The analysis of bond statistics further reveals that the crosslink elastic constant $K_{CL}$ plays a dual role. Increasing $K_{CL}$ enhances the elastic stiffness of individual bonds and the bonds sensitivity to deformation. This leads to a non-trivial redistribution of bond populations, with stiff bonds preferentially stabilizing polymer-polymer connections while destabilizing polymer-bacteria links due to geometric incompatibilities and local constraints. These results highlight that bond mechanics, kinetics, and network topology are deeply coupled in determining the structure of the system.

Finally, we have shown that this crossover can be quantitatively captured by a simple phenomenological model in which the fractions of each bond type emerge from a competition between effective statistical weights. These weights encode both energetic contributions, through the binding energies $E_{e,pp}$ and $E_{e,pb}$, and entropic/geometric factors associated with the availability of compatible partners. Within this framework, the crossover point $r^*$ is not determined solely by energetic differences, but by a balance between binding affinity and the progressive exhaustion of available bonding sites. This provides a direct and intuitive interpretation of the simulation results, and establishes a bridge between microscopic parameters and macroscopic network structure.

\textcolor{black}{Overall, our results demonstrate that the biofilm network structure can be finely tuned by independently controlling $E_{e,pp}$ and $E_{e,pb}$.  Rather than simply increasing or decreasing the total number of cross-links, energy asymmetry acts as a selective mechanism that determines which network dominates.  This provides a powerful route to control the microscopic organization of the biofilm, and anticipates significant consequences for its macroscopic mechanical properties. Work is in progress to unravel the stress response of this novel mesoscale biofilm model to an  applied  oscillatory strain.  } 

\section*{Acknowledgements}
J.M.R. acknowledges funding from Juan de la Cierva postdoctoral fellowship funding from Spanish Ministry of Education (JDC2024-053228-I). C.V. acknowledges funding from IHRC22/00002 and Proyecto PID2022-140407NB-C21 by MCIN/AEI/10.13039/501100011033 and FEDER, UE.

\section*{Data availability}
The data that support the findings of this study are available from the corresponding author upon reasonable request.

\section*{Declaration of competing interest}
The authors declare that they have no known competing financial interests or personal relationships that could have appeared to influence the work reported in this paper.

\bibliography{Biblio}

\newpage

\appendix

\section{Maximum number of links \label{append:A}} 

In the system under study, two types of structures are present: bacterias and polymers. The bacterias are composed of $B_b=400$ beads each, while the polymers consist of $B_p=100$ beads. A percent of the beads in each bacteria act as linkers, $r(\%)$, meaning they are capable of forming bonds with polymers. In contrast, all beads in the polymers function as linkers and are thus able to form bonds both with other polymers and with the linker beads of the bacterias. However, they cannot interact with the $100-r(\%)$ of bacteria's beads that are not linkers. An important constraint is that polymer beads cannot form intramolecular bonds—that is, they cannot bond with other beads within the same polymer chain.\\

\noindent
Given these restrictions, it is possible to define an upper bound for the total number of bonds that can be formed in an optimal configuration. This maximum is determined by the total number of available linker beads across both species and the limitations imposed by the bonding rules mentioned above. If the system is composed by $N_b$ bacterias and $N_p$ polymers, the total linker beads of the system are
\begin{equation}
L_b = N_b \cdot r \cdot B_b + N_p \cdot B_p 
\end{equation}
Since each bond connects two linker beads, the maximum number of bonds is given by
\[
N_{\text{max}} = \left\lfloor \frac{L_b}{2} \right\rfloor
\]
This is the theoretical upper bound, corresponding to the ideal case where all linker beads participate in bonding, no intramolecular bonds occur within polymer chains and each bond connects two distinct linker beads from different molecules.

\section{Monte Carlo time \label{append:tauMC}}

\noindent
As a first test we have the complete system with number of bacteria $N_b=184$, being the number of beads per polymer $B_b=440$; number of polymer $N_p=80$, with length $l_p=50 \, \sigma$, being the number of beads per polymer $B_p=100$, as the radius of the polymer beads is $\sigma_b=0.5\, \sigma$; number of solvent particles $N_s=35.000$, polymer-bacteria bonding energy $E_{e,f-p}=4$, polymer-polymer bonding energy $E_{e,p-p}=4$. Using this values for the system considering $r=1$, the maximum number of cross-links (CL) applying the equation found in previous section is $N_\text{max}=8000$.  \\

\noindent
The application time of the creation and destruction algorithm is vary, $\tau_{MC}=0.5 \, \tau_0, 2 \tau_0 \,$ and $5 \tau_0$ conserving in all simulations the number of time that MC algorithm is applied, so number of MC 'events' is $N_G=10^4$. The fraction of particles belonging to the bacteria considered as linkers in this first test was $r=0\%$, $20\%$ and $70\%$, while 100\% of the polymer beads were considered as linkers.\\

\begin{figure}[h!]
    \centering
    \includegraphics[width=0.87\linewidth]{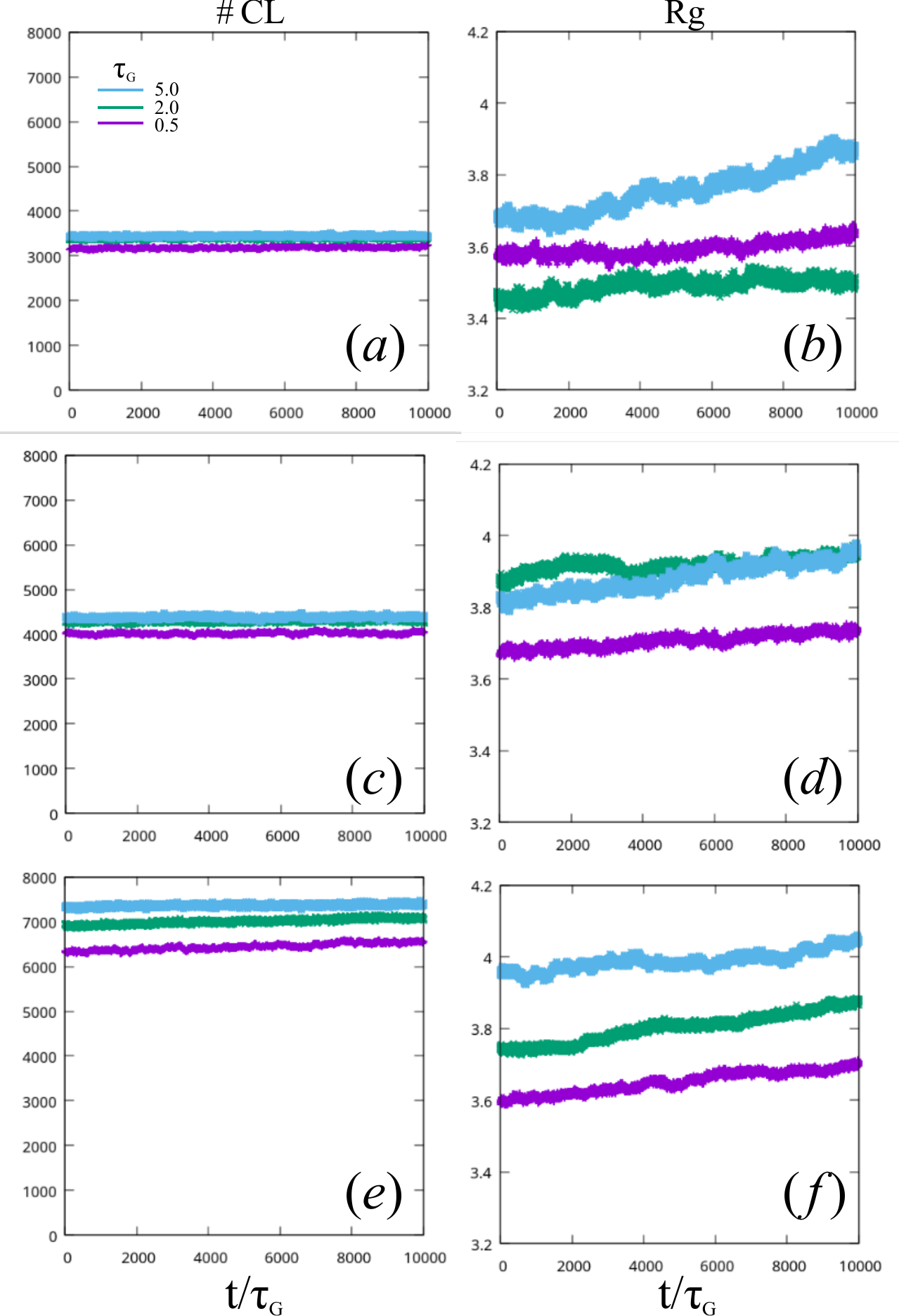}
    \caption{Number of Cross-Links (Left) and radius of gyration (Right) evolution for different values of the application time of the creation and destruction algorithm $\tau_{G}=0.5\tau_0,  \; 2 \tau_0, \;  5\tau_0$ and different values of $r=$ (a) 0.0, (b) 0.2 and (c) 0.7. Note that time is normalized by $\tau_G$ in such way all simulations have run over the same number of events in MC algorithm.}
    \label{fig:Sfig3}
\end{figure}

\noindent
To measure the steady state of the system we check the evolution of the total number of CL in the system (CL for each specie have been check but similar behavior was found, so data are not shown) and the radius of gyration of the polymers
\begin{equation}
    \text{R}_\text{g}(t) = \sqrt{\langle R_g^2 \rangle} =  \left[  \frac{1}{M} \sum_{j=1}^{B_p N_p} m_j \, \left( \textbf{r}_j-\textbf{r}_{cm} \right)^2 \right]^{1/2}
\end{equation}
where $M$ is the total mass of the polymers, $\textbf{r}_{cm}$ is the center of mass of all the polymers and the sum is over all the polymer beads in the system.\\

\noindent
As show in Fig. \ref{fig:Sfig3}, the independence of the steady state as a function of the MC time looking at the number of CL. A value of $\tau_{MC} = 2\tau_0$ is chosen for the results in the main text as it is sufficiently large to ensure that bonds are not created and destroyed too rapidly, while not being so large as to significantly slow down the system’s relaxation to the steady state. Note that even if $\text{R}_{\text{g}}$ is fluctuation and looks like it has not reach completely the steady state in this test simulations for the main text have run much longer and the value is evolving slowly in a range lower than the typical size of the beads of the polymer.  \\

\section{Activation energy sweep}

\begin{figure}[h!]
    \centering
    \includegraphics[width=0.87\linewidth]{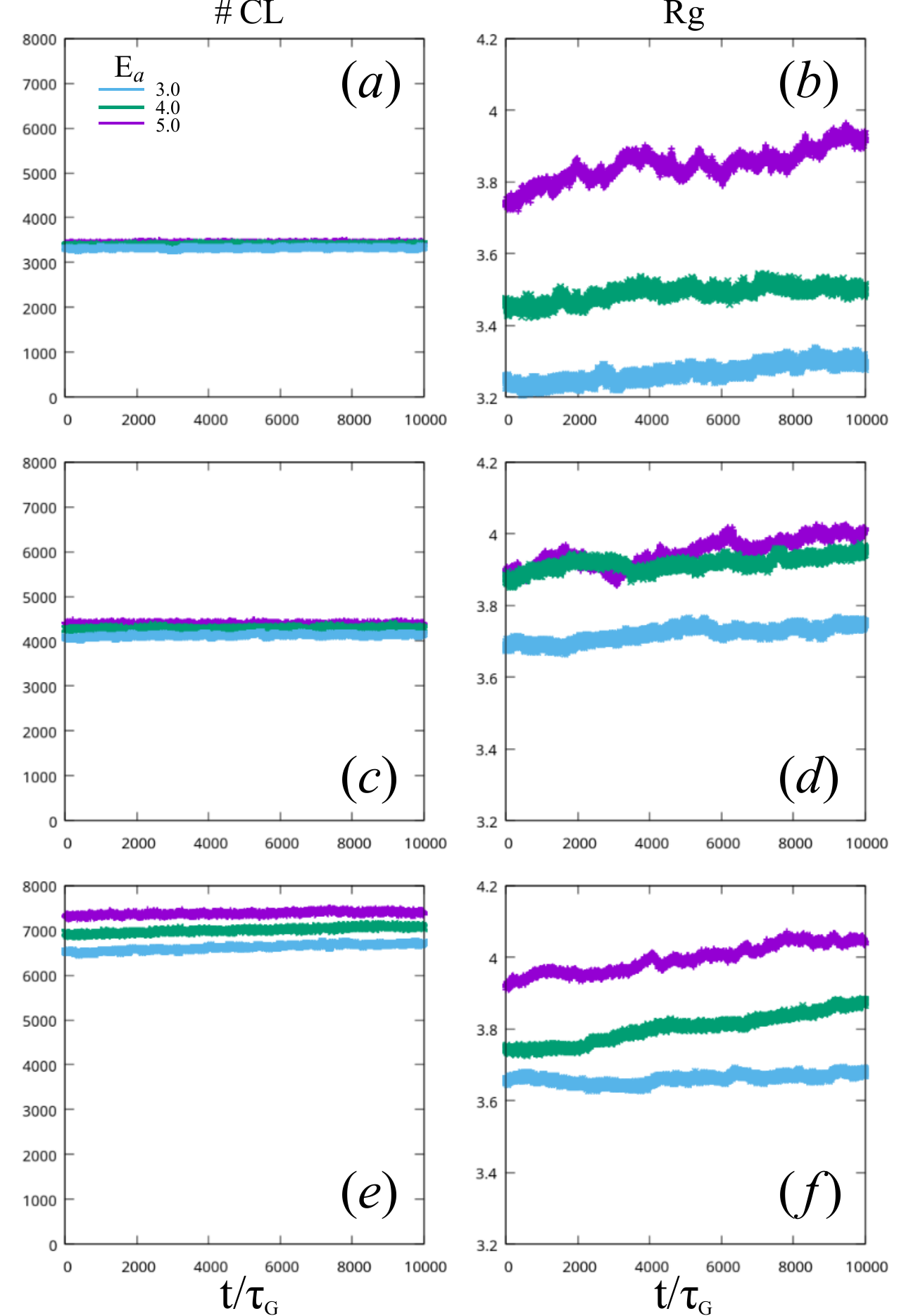}
    \caption{Number of Cross-Links (Left) and radius of gyration (Right) evolution for different values of activation energy $E_{a}=3.0,  \; 4.0 , \;  5.0$ and different values of $r=$ (a) 0.0, (b) 0.2 and (c) 0.7. Note that time is normalized by $\tau_G=2\tau_0$ in such way all simulations have run over the same number of events in MC algorithm.}
    \label{fig:Sfig2}
\end{figure}

\noindent
In this section we use the same parameters as in the previous one, except for the fact that we set $\tau_{MC}=2 \, \tau_0$ and change the activation energy of both species for three different values $E_{a,pp}=E_{a,pb}=3,$ 4 and 5. In Fig. \ref{fig:Sfig2}e show that the different activation energy leads similar steady-state in term of the number of CL, concluding a partial independency of the steady state in the choose of $E_a$ value for the simulations. On the other hand, the structure of the polymers, monitored by the radius of gyration, $\text{R}_\text{g}$, seems to be more sensitive on the activation energy. However, as happens in the previous case for the choose of $\tau_G$, the difference only appears in a range that oscillate about one bead diameter, that could be consider not relevant for this study.

\section{Statistical interpretation of the two-channel competition model}

\noindent
In order to rationalize the phenomenological expression used to describe the fraction of cross-links of each type, we formulate the problem as a competition between two alternative bonding states available to a generic polymer binding site. State $A$: the site forms a polymer-polymer (p-p) bond; and State $B$: the site forms a polymer--bacteria (p-b) bond.  At equilibrium, the probability of each state is determined by its statistical weight.  Within a standard statistical mechanics framework, these probabilities can be written as
\begin{equation}
P_A = \frac{e^{-F_A/RT}}{e^{-F_A/RT} + e^{-F_B/RT}}, 
\qquad
P_B = \frac{e^{-F_B/RT}}{e^{-F_A/RT} + e^{-F_B/RT}},
\label{eq:softmax}
\end{equation}
where $F_A$ and $F_B$ are the effective free energies associated with each bonding channel.  This expression corresponds to a two-state Boltzmann distribution and is mathematically equivalent to a softmax (logit) model. Identifying $P_A \equiv x$ and $P_B \equiv y$, Eq.~(\ref{eq:softmax}) provides a natural probabilistic interpretation of the fractions of cross-links.\\

\noindent
Each bonding state is characterized by a binding energy $E_A = -E_{e,pp}$ and $E_B = -E_{e,pb}$, so that stronger bonds correspond to lower energies.  These contributions favor the formation of the corresponding cross-link type. In addition to energetic contributions, the formation of a bond depends on the number of available partners.  This introduces an entropic term associated with configurational degeneracy.\\

\noindent
For polymer-polymer bonds, the availability of partners decreases as the fraction of active bacterial sites $r$ increases, since more polymer sites become engaged in p-b bonding.  To leading order, this effect can be represented as a factor proportional to $(1-r)$. Similarly, the availability of polymer-bacteria bonding configurations increases with $r$, since more bacterial sites are active, leading to a factor proportional to $r$.\\

\noindent
More generally, these effects can be expressed as effective degeneracy factors
\begin{equation}
\Omega_A \propto (1-r)^{\alpha}, 
\qquad 
\Omega_B \propto r^{\beta},
\end{equation}
where the exponents $\alpha$ and $\beta$ account for many-body effects, steric constraints, and spatial correlations beyond a simple mean-field description. Combining energetic and entropic contributions, the free energy of each state can be written as
\begin{equation}
F_A = -E_{e,pp} - RT \, \alpha \ln(1-r),
\end{equation}
\begin{equation}
F_B = -E_{e,pb} - RT \, \beta \ln r.
\end{equation}

\noindent
Introducing these free energies into Eq.~(\ref{eq:softmax}), we obtain
\begin{equation}
x = \frac{e^{(E_{e,pp}/RT)} (1-r)^{\alpha}}{e^{(E_{e,pp}/RT)} (1-r)^{\alpha} + e^{(E_{e,pb}/RT)} r^{\beta}},
\end{equation}
\begin{equation}
y = \frac{e^{(E_{e,pb}/RT)} r^{\beta}}{e^{(E_{e,pp}/RT)} (1-r)^{\alpha} + e^{(E_{e,pb}/RT)} r^{\beta}}.
\end{equation}

\noindent
Defining the statistical weights
\begin{equation}
w_{p-p} = e^{E_{e,pp}/RT}, 
\qquad 
w_{p-f} = e^{E_{e,pb}/RT},
\end{equation}
we recover the phenomenological expression used in the main text.\\

\noindent
This derivation shows that the proposed model is equivalent to a two-state statistical competition in which each bonding channel is weighted by a combination of energetic and entropic contributions.  The binding energies $E_{e,pp}$ and $E_{e,pb}$ control the intrinsic affinity of each bond type, while the factors $(1-r)^{\alpha}$ and $r^{\beta}$ encode the effective number of accessible configurations.\\

\noindent
Within this interpretation, the exponents $\alpha$ and $\beta$ acquire a clear physical meaning: they quantify how rapidly the configurational space available to each bonding channel changes with $r$, incorporating the effects of steric interactions, network constraints, and spatial correlations.

\vspace{0.5cm}

\noindent
We fix $r=0.7$, corresponding to a regime dominated by polymer--bacteria interactions but still sensitive to bond redistribution, and we vary the crosslink stiffness $K_{CL}=3, 30, 300$. A first important observation is that the mechanical response depends not only on the equilibrium number of CLs, but also on their dynamical stability. As discussed in Section~\ref{sec:gilespi}, the rupture propensity depends exponentially on the elastic energy stored in the bond. Therefore, increasing $K_{CL}$ makes bonds significantly more sensitive to small deviations from their equilibrium length. Under deformation, even moderate displacements can induce large elastic penalties, effectively lowering the activation barrier for rupture. As a consequence, for very large $K_{CL}$ (e.g. $K_{CL}=300$), bonds become highly fragile under shear, leading to enhanced bond breaking during deformation.

\noindent
In particular, increasing $K_{CL}$ leads to a stiffer but more fragile network: although individual bonds store more elastic energy, they are also more prone to rupture under deformation. This competition gives rise to a non-trivial dependence of the viscoelastic moduli on $K_{CL}$, and highlights the importance of considering bond dynamics explicitly when modeling biofilm mechanics.

\end{document}